\documentclass[11pt]{article}

\usepackage{a4,amsmath,amssymb,here,epsf,bbold}

\author{ Nico Stollenwerk$^{\# }$  \&  Ma\'ira Aguiar$^+ $  \\ 
\\{\small  
Centro de Matem\'atica e Aplica\c{c}\~oes Fundamentais 
da Universidade de Lisboa,}
\\{\small 
Avenida Prof. Gama Pinto 2, 1649-003 Lisboa, Portugal,}
\\{\small Instituto Gulbenkian de Ci\^encias, }
\\{\small  Apartado 14, Edificio Amerigo Vespucci, 2781-901 Oeiras, Portugal,} 
\\{\small Funda\c{c}\~ao Ezequiel Dias, 
Laborat\'orio de dengue e febre amarela,} 
\\{\small Rua Conde Pereira Carneiro 80, 30510-010 Belo Horizonte-MG, Brasil,}
\\{\small and Research Center J\"ulich, D-52425 J\"ulich, Germany}
\\{\footnotesize e-mails:  $^+$ {\tt maira@igc.gulbenkian.pt} 
and $^{\# }$ 
{\tt nks22@cam.ac.uk}  }
\\
}
\title{The SIRI stochastic model \\
		with creation and annihilation operators
}
\date{{\small \today } }

\begin{document}
\maketitle
\vspace{0.0cm}

\begin{abstract}
\noindent
We generalize the well known formulation of the susceptibles, infected,
susceptibles (SIS) spatial epidemics
with creation and annihilation operators to the reinfection model
including recovered which can be reinfected, the
SIRI model, using ladder operators constructed from the Gell-Mann matrices
known in quantum chromodynamics.
\end{abstract}


\section{Introduction}


Stochastic epidemic models like the susceptibles, infected,
susceptibles (SIS) spatial epidemics, also know as contact process,
are generalized to include more realistic scenarios,
e.g. including resistance and eventually milder reinfection,
giving the SIRI stochastic model. An analysis in terms of
pair approximation has been recently given for the SIRI model
to obtain phase transition lines
\cite{StollenwerkMartinsPinto2007}. This improves the rough 
qualitative analysis using mean field approximation, leading to the
often studied simple ordinary differential equation systems.
To obtain more insight into such stochastic epidemic models,
strating from the master equation formulation \cite{vanKampen}
a field theoretic formulation with creation and annihilation operators
giving a Schr\"odinger like operator equation has been used, first in
\cite{Doi}, then subsequently elaborated 
\cite{GrassbergerdelaTorre,GrassbergerScheunert,Peliti}
and more recently used 
e.g. in \cite{BrunelOerdingWijland,ParkPark2005} 
and e.g. in \cite{Oliveira2006}.

In the analysis of such stochastic spatial models
Pauli matrices are used to construct spin 1/2 ladder operators
which serve as creation and annihilation operators for a
formulation of the master equation in vector notation 
\cite{Hinrichsen2000} for the SIS epidemic model.
  We generalize to the SIRI epidemic model, finding ladder operators
as creation and annihilation operators derived from the
Gell-Mann matrices, normally used in quantum chromodynamics
(see e.g. \cite{Aguado2008}).

While mostly a subsequent renormlaization analysis based on
path integrals is performed \cite{BrunelOerdingWijland,ParkPark2005},
so that classical methods from particle physics can be used
\cite{ZinnJustin},
another way of series expansions based on a perturbation ansatz,
as described in 
\cite{DickmanJensen1991,Oliveira2006} is a promissing way to
obtain e.g. phase separation lines in the SIRI model. Up to now
only simple systems like the one dimensional contact process, i.e.
the SIS epidemics, have been analysed in this way.


\section{The SIS epidemic model revisited}


The Pauli matrices give lowering and raising operators for spin 1/2 systems
(named after Wolfgang Pauli, see  
\cite{ZinnJustin}).
The Pauli matrices are
\begin{equation}
	\sigma_x :=
	\left(
	\begin{array}{cc}
	0 & 1 \\
	1 & 0 \\
	\end{array}
	\right)
		\quad , \quad
	\sigma_y :=
	\left(
	\begin{array}{cc}
	0 & -i \\
	i & 0 \\
	\end{array}
	\right)
		\quad , \quad
	\sigma_z :=
	\left(
	\begin{array}{cc}
	1 & 0 \\
	0 & -1 \\
	\end{array}
	\right)
\label{paulimatrices}
\end{equation}
and from these the ladder operators \cite{BrunelOerdingWijland}
are:
the raising operator
\begin{equation}
	\sigma_+ := \frac{1}{2} \left( \sigma_x +i \sigma_y  \right) =
	\left(
	\begin{array}{cc}
	0 & 1 \\
	0 & 0 \\
	\end{array}
	\right)
\end{equation}
and the lowering operator
\begin{equation}
	\sigma_- := \frac{1}{2} \left( \sigma_x -i \sigma_y  \right) =
	\left(
	\begin{array}{cc}
	0 & 0 \\
	1 & 0 \\
	\end{array}
	\right)
		\quad.
\end{equation}
Hence for the states
\begin{equation}
	| 1 \rangle :=
	\left(
	\begin{array}{c}
	1 \\
	0 \\
	\end{array}
	\right)
		\quad , \quad
	| 0 \rangle :=
	\left(
	\begin{array}{c}
	0 \\
	1 \\
	\end{array}
	\right)
		\quad ,
\end{equation}
we have creation operator $c^+=\sigma _+ $ giving 
\begin{equation}
      c^+ | 0 \rangle = | 1 \rangle 
\end{equation}
and annihilation operator $c=\sigma _- $ giving
\begin{equation}
      c | 1 \rangle = | 0 \rangle 
      \quad.
\end{equation}
Then the SIS epidemic with master equation
\begin{eqnarray}
{\frac{d }{d t }} p(I_1, ... , I_N, t)
& = & 
  \sum_{i=1 }^{N}
	\;w_{I_i,1-I_i} (t) \;\; p(I_1, ... , 1-I_i, ... , I_N, t) 
					\nonumber 
\\
		& & \label{masterSIS}		
\\
	&  &
	- \sum_{i=1 }^{N}
	\;w_{1-I_i,I_i} (t) \;\; p(I_1, ... , I_i, ... , I_N, t)
				\nonumber 	
\end{eqnarray}
for $I_i \in \{0,1\}$ and 
transition rate
\begin{equation}
w_{I_i,1-I_i}  = \beta 	\left(
	\sum_{j=1 }^{N} \; J_{ij} I_j
				\right) \cdot I_i
	+\alpha \cdot (1-I_i)
	\quad,
\end{equation}
and 
\begin{equation}
w_{1-I_i,I_i}  = \beta 	\left(
	\sum_{j=1 }^{N} \; J_{ij} I_j
				\right) \cdot (1-I_i)
	+\alpha \cdot I_i
	\quad,
\end{equation}
can be given in vector notation by
\begin{equation}
{\frac{d }{d t }} | \Psi (t) \rangle  = L | \Psi (t) \rangle
	\label{sg1}	
\end{equation}
for a Liouville operator $L$ to be calculated from the master equation
(Eq. (\ref{masterSIS})) and
with state vector $| \Psi (t) \rangle  $ defined by
\begin{eqnarray}
 	| \Psi (t) \rangle 
	& := & 
 	 	\sum_{I_1=0 }^{1 } ... \sum_{I_N=0 }^{1 }
		p(I_1, ..., I_N,t) \left( c_1^+ \right) ^{I_1} ...  
                \left( c_N^+ \right) ^{I_N} 
		| 0 \rangle
		\label{state}			\nonumber 
\\
		& & \label{state1}		
\\
	&  =:&
		\sum_{ \{ I \} }^{} p(\{ I \} , t) 
			\left(
			\prod_{i=1}^{N} \left( c_i^+ \right) ^{I_i}
		 	\right)
		| 0 \rangle
		\label{state2}  		\nonumber 	
\end{eqnarray}
and vacuum state $| 0 \rangle $.

Parameters of the model are (for birth-death process or SIS epidemic)
$\beta $ birth or infection rate and $\alpha $ death or recovery rate.
Here $(J_{ij})$ is the adjacency matrix containing $0$ for no
connection and $1$ for a connection between sites $i$ and $j$,
hence $J_{ij}=J_{ji}\in \{0,1\}$ for $i\neq j$ and $J_{ii}=0$.

For lattice sites $i$ we have
creation and annihilation operators defined  by
$c_i^+ | 0 \rangle = | 1 \rangle $ and $c_i | 1 \rangle =  | 0 \rangle $,
and $\left( c_i^+ \right) ^2 | 0 \rangle = 0$ and 
$c_i | 0 \rangle =  0 $, hence
\begin{eqnarray}
	c_i \;\; | I_i \rangle & =& I_i \cdot |1-I_i \rangle \\
	c_i^+ | I_i \rangle & =& (1-I_i) \cdot |1-I_i \rangle 
\end{eqnarray}
and $\left( c_i^+ \right) ^2 | I_i \rangle = c_i^2 | I_i \rangle =0$.
We have anti-commutator rules on single lattice sites 
\begin{equation} 
	[c_i,c_i^+]_+ := c_ic_i^++c_i^+c_i =1
\end{equation}
and ordinary commutators for different lattice sites $i\neq j$
\begin{equation} 
	[c_i,c^+_j]_- := c_ic_j^+-c_j^+c_i =0
\end{equation}
respectively
\begin{equation} 
	[c_i,c_j]_- = 0 \quad , \quad  [c_i^+,c_j^+]_- =0
	\quad.
\end{equation}
These are exactly the raising and lowering operators as given
explicitly in 
\cite{BrunelOerdingWijland}.

The dynamics is expressed by
\begin{equation}
{\frac{d }{d t }} | \Psi (t) \rangle  
=
\sum_{ \{ I \} }^{} 
\left( {\frac{d }{d t }} p(\{ I \} , t) \right)
\prod_{i=1}^{N} \left( c_i^+ \right) ^{I_i}
	 | 0 \rangle
= L | \Psi (t) \rangle
\end{equation}
where the master equation has to be inserted and evaluated to obtain the
specific form of the operator $L$.
The Liouville operator is after some calculation
\begin{equation}
L = \sum_{i=1 }^{N} (1- c_i)  \beta 
	\left(
		 \sum_{j=1 }^{N} J_{ij} c^+_j c_j
	\right) c^+_i
	+\sum_{i=1 }^{N} (1- c_i^+) \alpha  \; c_i
		\quad.
	\label{liouville}
\end{equation}
The term 
$(1- c_i)$ guarantees the normalization of the master equation solution
and $\beta  J_{ij} c^+_j c_j c^+_i $ creates one infected at site $i$ from a
neighbour $j$ which is itself not altered. $c^+_j c_j$ is simply the number
operator on site $j$.
Furtheron, $\alpha   c_i$ removes a particle from site $i$, again ensuring
normalization with $(1- c_i^+)$. 
This form Eq. (\ref{sg1}) and Eq. (\ref{liouville}) 
is exactly the form given as well
in  \cite{GrassbergerdelaTorre}, 
there pp. 392--394,
Appendix A), hence using the raising and lowering operators.

Mean values for any quantity can now be given with a Felderhof projection
state \cite{Felderhof}.


\subsection{Perturbation analysis gives critical threshold}


The Liouville operator can be given in the form  of a perurbation
ansatz with an easily diagonalizable free operator $W_0 $, acting
only on single sites,
and
an interaction operator $V $ contributing with strength $\lambda $
to the interaction (see  for the 1 dimensional contact process  e.g.
\cite{Oliveira2006})
\begin{equation}
L = W_0 + \lambda \cdot V
	\label{liouvilleperturbation}
\end{equation}
with
\begin{equation}
W_0 := \sum_{i=1 }^{N} \hat B_i 
\end{equation}
with $\hat B_i := (1- c_i^+)\; c_i$ and 
without loss of generality $\alpha =1 $,
and with
\begin{equation}
V := \sum_{i=1 }^{N} \hat Q_i (\hat n_{i-1} + \hat n_{i+1} )
\end{equation}
where $\hat Q_i :=\frac{1}{Q}(1- c_i) \; c^+_i$ 
with $Q=2$ the number of neighbours in two dimensions, 
and finally originating from
the interaction term in one dimension
\begin{equation}
 \sum_{j=1 }^{N} J_{ij} c^+_j c_j = c^+_{i-1} c_{i-1} + c^+_{i+1} c_{i+1}
 \quad.
\end{equation}
Identifying 
\begin{equation}
\lambda := \beta \cdot Q
\end{equation}
completes the perturbation ansatz Eq. (\ref{liouvilleperturbation})
for the Liouville operator given in Eq. (\ref{liouville}), as given in
\cite{Oliveira2006}. With this ansatz the critical threshold and
the critical exponents can be calculated very accurately via
a scaling argument, e.g. for the time correlation
function using the spectral gap \cite{Oliveira2006}, 
and Pad\'e approximation.


\section{Generalization to the SIRI epidemic model}


To generalize from the SIS epidemic model to more general cases 
like the SIRI we need to extend the number of dimensions of the
basic vectors for the states
(see e.g. \cite{Hinrichsen2000}, p. 132), hence
one more dimension for another particle type $R$ besides $I$.
For SIS we have
\begin{equation}
	| 0 \rangle :=
	\left(
	\begin{array}{c}
	0 \\
	1 \\
	\end{array}
	\right) =: | S \rangle
		\quad , \quad
	| 1 \rangle :=
	\left(
	\begin{array}{c}
	1 \\
	0 \\
	\end{array}
	\right) =: | I \rangle
\end{equation}
For the SIRI model hence we generalize to
\begin{equation}
	| S \rangle :=
	\left(
	\begin{array}{c}
      	0 \\
	0 \\
	1 \\
	\end{array}
	\right)
		\quad , \quad
	| I \rangle :=
	\left(
	\begin{array}{c}
      	0 \\
	1 \\
	0 \\
	\end{array}
	\right) 
		\quad , \quad
	| R \rangle :=
	\left(
	\begin{array}{c}
      	1 \\
	0 \\
	0 \\
	\end{array}
	\right) 
        \quad.
\end{equation}
Then creation and annihilation operators are given by
\begin{equation}
      a^+ | S \rangle = | I \rangle 
      		\quad , \quad
      a | I \rangle = | S \rangle
\end{equation}
for infected, and for recovered we have
\begin{equation}
      b^+ | S \rangle = | R \rangle 
      		\quad , \quad
      b | R \rangle = | S \rangle
\end{equation}
being achieved by the following matrix representation
\begin{equation}
	a^+ :=
	\left(
	\begin{array}{ccc}
	0 & 0 & 0 \\
	0 & 0 & 1 \\
        0 & 0 & 0 \\
	\end{array}
	\right)
		\quad , \quad
	a :=
	\left(
	\begin{array}{ccc}
	0 & 0 & 0 \\
	0 & 0 & 0 \\
        0 & 1 & 0 \\
	\end{array}
	\right)
\end{equation}
and 
\begin{equation}
	b^+ :=
	\left(
	\begin{array}{ccc}
	0 & 0 & 1 \\
	0 & 0 & 0 \\
        0 & 0 & 0 \\
	\end{array}
	\right)
		\quad , \quad
	b :=
	\left(
	\begin{array}{ccc}
	0 & 0 & 0 \\
	0 & 0 & 0 \\
        1 & 0 & 0 \\
	\end{array}
	\right)
        \quad.
\end{equation}
We can include the old two dimensional creation and annihilation
operators, now extended to three dimensions, as
\begin{equation}
	c^+ :=
	\left(
	\begin{array}{ccc}
	0 & 1 & 0 \\
	0 & 0 & 0 \\
        0 & 0 & 0 \\
	\end{array}
	\right) = b^+ a
		\quad , \quad
	c :=
	\left(
	\begin{array}{ccc}
	0 & 0 & 0 \\
	1 & 0 & 0 \\
        0 & 0 & 0 \\
	\end{array}
	\right) = a^+ b
\end{equation}
which can be expressed in terms of $a^+$, $a$ and $b^+$, $b$. They give
\begin{equation}
      c^+ | I \rangle = | R \rangle 
      		\quad , \quad
      c | R \rangle = | I \rangle
      \quad.
\end{equation}
These operators $a^+ $, $a$, $b^+$, $b$ and $c^+$, $c$ are the ladder operators
of the 8 Gell-Mann matrices $\lambda _1 $, $\lambda _2 $, ... , $\lambda _8 $
\begin{equation}
      a^+ = \frac{1}{2} \left( \lambda _6 + i \lambda _7 \right)
      		\quad , \quad
      a = \frac{1}{2} \left( \lambda _6 - i \lambda _7 \right)
      \quad,
\end{equation}
\begin{equation}
      b^+ = \frac{1}{2} \left( \lambda _4 + i \lambda _5 \right)
      		\quad , \quad
      b = \frac{1}{2} \left( \lambda _4 - i \lambda _5 \right)
      \quad,
\end{equation}
\begin{equation}
      c^+ = \frac{1}{2} \left( \lambda _1 + i \lambda _2 \right)
      		\quad , \quad
      c = \frac{1}{2} \left( \lambda _1 - i \lambda _2 \right)
      \quad,
\end{equation}
with the Gell-Mann matrices
(named after Murray Gell-Mann, a generalization to the Pauli matrices),
well known in quatum chromodynamics (QCD) as representations of the
SU(3) group (special unitary group of dimension 3)
\begin{equation}
	\lambda _1=
	\left(
	\begin{array}{ccc}
	0 & 1 & 0 \\
	1 & 0 & 0 \\
        0 & 0 & 0 \\
	\end{array}
	\right)
		\quad , \quad
	\lambda _2=
	\left(
	\begin{array}{ccc}
	0 & -i & 0 \\
	i & 0 & 0 \\
        0 & 0 & 0 \\
	\end{array}
	\right)
\end{equation}
\begin{equation}
	\lambda _4=
	\left(
	\begin{array}{ccc}
	0 & 0 & 1 \\
	0 & 0 & 0 \\
        1 & 0 & 0 \\
	\end{array}
	\right)
		\quad , \quad
	\lambda _5=
	\left(
	\begin{array}{ccc}
	0 & 0 & -i \\
	0 & 0 & 0 \\
        i & 0 & 0 \\
	\end{array}
	\right)
\end{equation}
\begin{equation}
	\lambda _6=
	\left(
	\begin{array}{ccc}
	0 & 0 & 0 \\
	0 & 0 & 1 \\
        0 & 1 & 0 \\
	\end{array}
	\right)
		\quad , \quad
	\lambda _7=
	\left(
	\begin{array}{ccc}
	0 & 0 & 0 \\
	0 & 0 & -i \\
        0 & i & 0 \\
	\end{array}
	\right)
\end{equation}
and 
\begin{equation}
	\lambda _3=
	\left(
	\begin{array}{ccc}
	1 & 0 & 0 \\
	0 & -1 & 0 \\
        0 & 0 & 0 \\
	\end{array}
	\right)
		\quad , \quad
	\lambda _8= \frac{1}{\sqrt{3}}
	\left(
	\begin{array}{ccc}
	1 & 0 & 0 \\
	0 & 1 & 0 \\
        0 & 0 & -2 \\
	\end{array}
	\right)
\end{equation}
applied usually in QCD to the color states
\begin{equation}
	| red \rangle =
	\left(
	\begin{array}{c}
      	1 \\
	0 \\
	0 \\
	\end{array}
	\right)
		\quad , \quad
	| green \rangle =
	\left(
	\begin{array}{c}
      	0 \\
	1 \\
	0 \\
	\end{array}
	\right) 
		\quad , \quad
	| blue \rangle =
	\left(
	\begin{array}{c}
      	0 \\
	0 \\
	1 \\
	\end{array}
	\right) 
        \quad.
\end{equation}
See e.g. \cite{Hiesmayr2006} and
\cite{KoniorczykJanszky2001}.

For the commutator rules we now have
\begin{equation}
     a \;  a^+ = {\mathbb 1} -(a^+ a + b^+ b)
      \quad,
\end{equation}
and 
\begin{equation}
     b \; b^+ = {\mathbb 1} -(a^+ a + b^+ b)
      \quad,
      \label{commutationbb+}
\end{equation}
see  \cite{ParkPark2005}.


\subsection{The SIRI epidemic model
\label{Sirimodel}}


We consider the following transitions between host classes for $N $ individuals
being either susceptible $S $, infected $I$ by a disease or recovered $R$
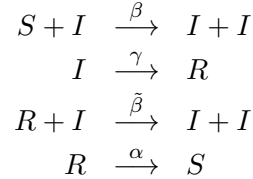
\begin{eqnarray}
	S+I	& \stackrel{\beta  }{\longrightarrow }  &  	I+I
	\nonumber
\\
	I 	& \stackrel{\gamma }{\longrightarrow }  &  		R
	\nonumber
\\
	R +I	& \stackrel{\tilde{\beta} }{\longrightarrow }  &  I+	I
	\nonumber
\\
	R	& \stackrel{\alpha }{\longrightarrow }  &  		S
	\nonumber
\end{eqnarray}
resulting in the master equation \cite{vanKampen}
for variables $S_i $, $I_i $ and $R_i \in \{ 0,1 \} $, $i=1,2,..., N $,
for $N$ individuals eventually on a regular grid,
with constraint $S_i+I_i+R_i=1 $.

The first infection $ S+I \stackrel{\beta  }{\longrightarrow }	I+I $
occurs with infection rate $\beta  $, whereas after recovery with rate
$\gamma $ the respective host becomes resistant up to a possible 
reinfection $R +I \stackrel{\tilde{\beta} }{\longrightarrow } I+I $
with reinfection rate $\tilde{\beta} $. Hence the recovered are only
partially immunized. For further analysis of possible stationary states
we include a transition from recovered to susceptibles $\alpha $, 
which might be
simply due to demographic effects (or very slow waning immunity for some
diseases). We
will later consider the limit of vanishing or very small $\alpha $. In case
of demography that would be in the order of inverse 70 years, whereas for 
the basic epidemic processes like first infection $\beta $ we would expect
inverse a few weeks.

The master equation is explicitly given in the 
following form, as e.g. described in \cite{StollenwerkMartinsPinto2007}
\begin{eqnarray}
{\frac{d }{d t }} & p  & (S_1,I_1,R_1,S_2,I_2,R_2,  ... , R_N, t)\nonumber
\\
& = & 
  \sum_{i=1 }^{N}   \beta 	\left(
	\sum_{j=1 }^{N} \; J_{ij} I_j \right) (1-S_i)
        \;\; p(S_1,I_1,R_1, ... ,1-S_i, 1-I_i,R_i ... , R_N, t) 
		\label{masterspacialsirialpha}		 \nonumber 
\\
		& &  \nonumber 
\\
	&  &
	+ \sum_{i=1 }^{N} \gamma (1-I_i)
        \;\; p(S_1,I_1,R_1, ... ,S_i, 1-I_i,1-R_i ... , R_N, t)
		 \nonumber 
\\
		& & 
\\
 	&  &
        + \sum_{i=1 }^{N}   \tilde{\beta} 	\left(
	\sum_{j=1 }^{N} \; J_{ij} I_j \right) (1-R_i)
        \;\; p(S_1,I_1,R_1, ... ,S_i, 1-I_i,1-R_i ... , R_N, t)
		 \nonumber 
\\
		& &  \nonumber 
\\
 	&  &
	+ \sum_{i=1 }^{N} \alpha (1-R_i)
        \;\; p(S_1,I_1,R_1, ... ,1-S_i, I_i,1-R_i ... , R_N, t)
		 \nonumber
\\
		& &  \nonumber 
\\
 	&  &
        - \sum_{i=1 }^{N}
        \left[
          \beta 	\left(
	    \sum_{j=1 }^{N} \; J_{ij} I_j \right) S_i
          + \gamma I_i
          + \tilde{\beta } 	\left(
	    \sum_{j=1 }^{N} \; J_{ij} I_j \right) R_i
          + \alpha R_i
          \right]		 
          \nonumber
\\
		& &  \nonumber 
\\
 	&  &
          \;\;\; 
          \quad \quad \quad \quad \quad \quad \quad \quad \quad
          \quad \quad \quad \quad \quad \quad \quad \quad \quad
          \quad \quad \quad \quad 
          \cdot  p(...S_i,I_i,R_i ...)
          \nonumber
	\quad.
\end{eqnarray}
$J_{i,j} \in \{ 0,1\}$ are the elements of the $N \times N $ 
adjacency matrix $J $, symmetric and with zero diagonal elements.
The formulation of the master equation is given in analogy to
the one used e.g. in \cite{Glauber}.


\subsection{Transitions in the SIRI model}


From the rules above we can obtain the different parts of the
Liouville operator corresponding to the transitions in the SIRI model
(derived from the master equation transition rates).

Explicitly we have
\begin{equation}
	S+I	 \stackrel{\beta  }{\longrightarrow }   	I+I
 \quad \quad, \quad \quad
a^+ | S \rangle  = | I \rangle
\end{equation}
means creation of an infected from a susceptible in interaction with
another infected,
hence 
\begin{equation}
L _{\beta } = \beta  \sum_{i=1 }^{N} (1- a_i) 
	\left(
		 \sum_{j=1 }^{N} J_{ij} a^+_j a_j
	\right) a^+_i
	\label{liouvillebeta}
\quad.
\end{equation}

Next we have
\begin{equation}
	R	 \stackrel{\alpha }{\longrightarrow }   		S
 \quad \quad, \quad \quad
b | R \rangle  = | S \rangle
\end{equation}
means creation of a susceptible from a recovered,
hence 
\begin{equation}
L _{\alpha } = \alpha  \sum_{i=1 }^{N} (1- b_i^+)  \; b_i
	\label{liouvillealpha}
\quad.
\end{equation}

Then we have
\begin{equation}
	I 	 \stackrel{\gamma }{\longrightarrow }   		R
 \quad \quad, \quad \quad
c^+ | I \rangle  = | R \rangle
\end{equation}
means creation of a recovered from an infected,
hence 
\begin{eqnarray}
L _{\gamma } & = & \gamma  \sum_{i=1 }^{N} (1- c_i)   \; c_i^+ 
	\nonumber
\\
             & = & \gamma  \sum_{i=1 }^{N} (1- a_i^+b_i)  \; b_i^+ a_i
	\label{liouvillegamma}
\end{eqnarray}

Finally, we have
\begin{equation}
	R +I	 \stackrel{\tilde{\beta} }{\longrightarrow }    I+	I
 \quad \quad, \quad \quad
c | R \rangle  = | I \rangle
\end{equation}
means creation of an infected from a recovered in interaction with
another infected,
hence 
\begin{eqnarray}
L _{\tilde{\beta} } & = & \tilde{ \beta } \sum_{i=1 }^{N} (1- c_i^+)  
	\left(
		 \sum_{j=1 }^{N} J_{ij} c_j c^+_j
	\right) c_i
	\nonumber
\\
             & = & \tilde{ \beta } \sum_{i=1 }^{N} (1- b_i^+ a_i)  
	\left(
		 \sum_{j=1 }^{N} J_{ij}    a_j^+b_j   b_j^+ a_j    
	\right) a_i^+b_i
	\label{liouvillebetatilde}
\end{eqnarray}
With the commutation rules, especially Eq. (\ref{commutationbb+}),
we obtain
\begin{equation}
      c \; c^+ = a^+ b \; b^+ a = a^+ ({\mathbb 1}-a^+ a - b^+ b) a
              = a^+ a
\end{equation}
the number operator for the infected, hence
\begin{eqnarray}
L _{\tilde{\beta} } 
             & = & \tilde{ \beta } \sum_{i=1 }^{N} (1- b_i^+ a_i)  
	\left(
		 \sum_{j=1 }^{N} J_{ij}    a_j^+ a_j    
	\right) a_i^+b_i
	\label{liouvillebetatilde2}
\end{eqnarray}

Then the total Liouville operator is given by the sum of the individual
Liouvilleans for the different transitions
\begin{equation}
      L= L_{\gamma } + L_{\alpha } + L _{\beta } + L _{\tilde{\beta} }
\end{equation}
where the first two terms $L_{\gamma } + L_{\alpha } $ are single site
or free contributions and the last two $L _{\beta } + L _{\tilde{\beta} } $
interaction contributions to the Liouville operator, suitable for
a perturbation analysis analoguous 
to ansatz Eq. (\ref{liouvilleperturbation}).


\section{Conclusions}


We have formulated the stochastic spatial reinfection model SIRI in
terms of creation and annihilation operators in a Schr\"odinger equation
like form, which opens the way for further analysis e.g. in terms of
series expansions via a perturbation ansatz to calculate critical thresholds
and critical exponents. The creation and annihilation operators
turn out to be the ladder operators of the Gell-Mann matrices of the
SU(3) group representation.


\section{Acknowledgments}  


We would like to thank Gabriela Gomes, Lisbon, 
Jos\'e Martins, Leir\'ia, Alberto Pinto, Braga, 
and Friedhelm Drepper, J\"ulich, for valuable discussions
on reinfection models, and Peter Grassberger, J\"ulich and Calgary, John Cardy,
Oxford, and Ronald Dickman, Belo Horizonte, for encouraging
discussions on the operator formulation of stochastic processes.





\begin{thebibliography}{99}




\bibitem[Stollenwerk, Martins and Pinto (2007)]
{StollenwerkMartinsPinto2007}
        Stollenwerk, N., Martins, J., \& Pinto, A. (2007)
        The phase transition lines in pair approximation for the basic 
        reinfection model SIRI, 
	{\it Physics Letters A} {\bf 371}, 379--388.


\vspace*{-0.2cm}


\bibitem[Park and Park (2005)]
{ParkPark2005}
        Park, S.C.,  \& Park, J.M. (2005)
        Generating function, path integral representation, and
        equivalence for stochastic exclusive particle systems,
        {\it Physical Review } {\bf E 71}, 026113.


\vspace*{-0.2cm}


\bibitem[de Oliveira (2006)]
{Oliveira2006}
        de Oliveira, M.J. (2006)
        Perturbation series expansion for the gap of the evolution operator
        associated with the contact process,
	{\it Physical Review } {\bf E 74}, 041121.


\vspace*{-0.2cm}

\bibitem[Grassberger and Scheunert (1980)]
{GrassbergerScheunert}
	Grassberger, P., \&  Scheunert, M. (1980) Fock-space methods for 
	identical classical objects,
	{\it Fortschritte der Physik} {\bf 28}, 547--578. 


\vspace*{-0.2cm}

\bibitem[Grassberger and de la Torre (1979)]
{GrassbergerdelaTorre}
	Grassberger, P., \& de la Torre, A. (1979)
	Reggeon Field Theory (Schl\"ogel's First Model) on a Lattice:
	Monte Carlo Calculations of Critical Behaviour,
	{\it Annals of Physics }{\bf 122}, 373--396.

\vspace*{-0.2cm}


\bibitem[van Kampen (1992)]
{vanKampen} 
	van Kampen, N.G. (1992) {\it Stochastic Processes in Physics 
	and Chemistry}
	(North-Holland, Amsterdam).


\vspace*{-0.2cm}

\bibitem[Zinn-Justin (1989)]
{ZinnJustin} 
	Zinn-Justin, J. (1989) {\it Quantum Field Theory and 
	critical phenomena}
	(Oxford University Press, Oxford).


\vspace*{-0.2cm}

\bibitem[Felderhof (1971)]
{Felderhof}
	Felderhof , B.U. (1971)  Spin relaxation of the Ising chain,
	{\it Rep. math. Phys.} {\bf 1}, 215--234.



\vspace*{-0.2cm}

\bibitem[Brunel, Oerding and van Wijland (2000)]
{BrunelOerdingWijland} 
	Brunel, V., Oerding, K., \&  Wijland, F. (2000) Fermionic 
	field theory for directed percolation in (1+1)-dimension, 
	{\it J. Phys. } {\bf A 33}, 1085--1097. 



\vspace*{-0.2cm}

\bibitem[Glauber (1963)]
{Glauber} 
	Glauber, R.J. (1963) Time-dependent statistics of the Ising model,  
	{\it J. Math. Phys.} {\bf 4}, 294--307. 


\vspace*{-0.2cm}

\bibitem[Ising (1925)]
{Ising}
	Ising, E. (1925)
	Beitrag zur Theorie des Ferromagnetismus,
	{\it Zeitschrift f\"ur Physik }{\bf 31}, 253--258.


\vspace*{-0.2cm}

\bibitem[Hinrichsen (2000)]
{Hinrichsen2000}
        Hinrichsen, H. (2000)
        Nonequilibrium critical phenomena and transition into absorbing states,
        {\it arxiv: cond-mat/0001070v2}
        (also available in {\it Advances in Physics}).

\vspace*{-0.2cm}

\bibitem[Aguado (2008)]
{Aguado2008}
        Aguado, M., Asorey, M., Ercolessi, E., Ortolani, F., \&
        Pasini, S. (2008)
        Numerical simulation of the SU(3) AFM Heisenberg model,
        {\it arxiv: 0801.3565v1}.



\vspace*{-0.2cm}

\bibitem[Hiesmayr (2006)]
{Hiesmayr2006}
        Hiesmayr, B.C., Koniorczyk, M., \& Narnhofer, H. (2006) 
        Maximizing nearest-neighbor entanglement in finitely correlated
        qubit chains,
	{\it Physical Review }{\bf A 73}, 032310(11).




\vspace*{-0.2cm}

\bibitem[Koniorczyk and Janszky (2001)]
{KoniorczykJanszky2001}
        Koniorczyk, M., \& Janszky, J. (2001)
        Photon number conservation and photon interference,
        {\it arxiv: quant-ph/0110170v2}.

\vspace*{-0.2cm}


\bibitem[Peliti (1985)]
{Peliti} 
	Peliti, L. (1985) Path integral approach to birth-death processes
	on a lattice,  
	{\it J. Physique} {\bf 46}, 1469--1483. 


\vspace*{-0.2cm}

\bibitem[Doi (1976)]
{Doi} 
	Doi, M. (1976) Stochastic theory of diffusion-controlled reactions,
	{\it J. Phys.} {\bf A 9}, 1479--1495. 


\vspace*{-0.2cm}

\bibitem[Dickman and Jensen (1991)]
{DickmanJensen1991} 
	Dickman, R.,  \& Jensen, I. (1991) 
        Time-Dependent perturbation theory for nonequilibrium lattice models,
	{\it Physical Review Letters} {\bf 67}, 2391--2394. 




\end{thebibliography}
\end{document}